\documentclass[journal]{vgtc}                     % final (journal style)
\onlineid{1102}

\vgtccategory{Research}

\vgtcpapertype{Design Studies}

\title{A Design Study Process Model for Medical Visualization}

\author{%
  Mengjie Fan and 
  Liang Zhou*
}

\authorfooter{
  \item
  	Mengjie Fan and Liang Zhou are with the Institute of Medical Technology, Peking University Health Science Center, and the National Institute of Health Data Science (NIHDS), Peking University.
    Liang Zhou is the corresponding author.
  	E-mails: mengjiefan@bjmu.edu.cn, zhoulng@pku.edu.cn.
}

\abstract{%
We introduce a design study process model for medical visualization based on the analysis of existing medical visualization
and visual analysis works, and our own interdisciplinary research experience. 
With a literature review of related works covering various data types and applications, we identify features of medical visualization and visual analysis research and formulate our model thereafter. 
Compared to previous design study process models, our new model emphasizes: distinguishing between different stakeholders and target users before initiating specific designs, distinguishing design stages according to analytic logic or cognitive habits, and classifying task types as inferential or descriptive, and further hypothesis-based or hypothesis-free based on whether they involve multiple subgroups. 
In addition, our model refines previous models according to the characteristics of medical problems and provides referable guidance for each step. 
These improvements make the visualization design targeted, generalizable, and operational, which can adapt to the complexity and diversity of medical problems. 
We apply this model to guide the design of a visual analysis method and reanalyze three medical visualization-related works.
These examples suggest that the new process model can provide a systematic theoretical framework and practical guidance for interdisciplinary medical visualization research. 
We give recommendations that future researchers can refer to, report on reflections on the model, and delineate it from existing models.
  
}


\graphicspath{{figs/}{figures/}{pictures/}{images/}{./}} % where to search for the images

\usepackage{tabu}                      % only used for the table example
\usepackage{booktabs}                  % only used for the table example
\usepackage{lipsum}                    % used to generate placeholder text
\usepackage{mwe}                       % used to generate placeholder figures

\usepackage{amsmath}
\usepackage{amsfonts}
\usepackage{enumitem}
\usepackage{algorithm}
\usepackage{algpseudocode}
\usepackage{mathptmx}                  % use matching math font
\usepackage{multirow}
\usepackage{color}

\newcommand{\rev}[1]{#1}
\usepackage{float}  % 支持[H]定位
\usepackage{tabularx}  % 自动调整列宽
\usepackage{ragged2e}  % 提供更好的对齐和换行控制
\usepackage{booktabs}       % 三线表
\usepackage{threeparttable} % 表格备注支持
\usepackage[utf8]{inputenc}

\usepackage{mathptmx}                  % use matching math font

\begin{document}

%%%%%%%%%%%%%%%%%%%%%%%%%%%%%%%%%%%%%%%%%%%%%%%%%%%%%%%%%%%%%%%%
%%%%%%%%%%%%%%%%%%%%%% START OF THE PAPER %%%%%%%%%%%%%%%%%%%%%%
%%%%%%%%%%%%%%%%%%%%%%%%%%%%%%%%%%%%%%%%%%%%%%%%%%%%%%%%%%%%%%%%

\firstsection{Introduction}

\maketitle

Medical visualization aims to help practitioners and researchers better understand and analyze medical data and information by transforming medical data into visual forms such as graphs and images through computer graphics and image processing techniques~\cite{Meuschke2022}.
Medical visualization is a key research area in the current data-driven healthcare practice, focusing on the acquisition of medical evidence and helping stakeholders to understand and analyze medical data. 
Visualization and visual analysis of medical data enables supporting clinical decision making, improving healthcare, simplifying presentation of healthcare data, and accelerating healthcare performance, etc.~\cite{Abudiyab2022, Rea2022}.

Design study is an increasingly popular form of problem-driven visualization research~\cite{Sedlmair2012}. 
Sedlmair et al. define a design study as an interdisciplinary research method in which visualization researchers analyze a specific real-world problem faced by domain experts and design a visualization method to support the solution of the problem~\cite{Sedlmair2012}.
Tasks in the real world can be classified into problem-driven tasks and technology-driven tasks~\cite{Keim2008}. 
Medical data analysis tasks are mostly problem-driven, and their main goals are to use various medical data to address specific problems in real medical scenarios.
The design study methodology can take advantage of the synergy between visualization and medical experts to create tailored visualization techniques and solutions for diverse needs, thereby addressing specific medical challenges.

Although some design study process models and practical guidelines are available to guide general visualization design practices, there is still a lack of a systematic methodology considering the specialty of the area of medicine to guide the specific design and research process in medical visualization. 
This absence often makes it difficult for researchers to realize the full potential of visualization when solving complex problems in medicine, and it also limits the efficiency of interdisciplinary collaboration. 

\begin{figure*}[htb]
    \centering
    \includegraphics[width=0.98\linewidth, trim={0cm 0.1cm 0cm 0.1cm},clip]{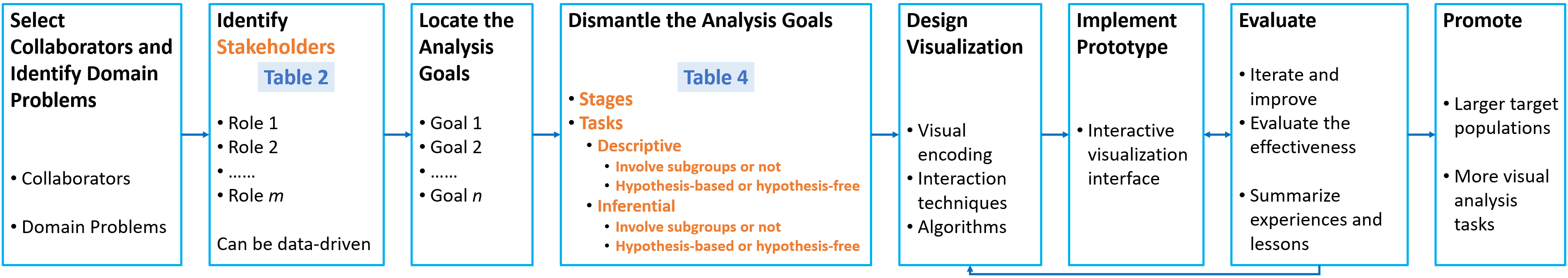}

    \caption{The design study process model applicable for medical visualization. Steps of the model are elaborated in~\cref{sec:process_model_whole}. New factors introduced in our model are in orange, and details thereof are summarized in~\cref{tab:stakeholders_target} and~\ref{tab:dismantle_goals}.} 
    \label{fig:process_model}
\end{figure*}

In this paper, we propose a design study process model for medical visualization and visual analysis.
The model is tailored for the characteristics of medical data and analysis tasks based on a literature review and our own experiences.
Our first contribution is that the specialized model covers the whole process from collaborator selection all the way to the final evaluation through steps of visualization solution implementation, as shown in~\cref{fig:process_model}.
We identify three factors that are critical for medical visualization, namely, \emph{stakeholders}, \emph{stages}, and \emph{subgroup analysis}.
At each stage of the model, we provide referable guidance to assist researchers and practitioners in finding an appropriate visualization solution that addresses a specific medical problem.
The usefulness of the model is demonstrated through four use cases of medical visualization works.

Our second contribution includes several recommendations for medical visualization design studies.
We recommend that researchers conduct thorough discussions with collaborators to ensure a proper and consistent understanding of the specific domain problem. 
In a specific visualization design, the relevant stakeholders and the final target users should be determined, different stages can be distinguished according to the analytic logic and cognitive habits of knowledge acquisition, and task types can be distinguished according to whether they involve subgroups. 
We recommend prioritizing a controlled study to evaluate the designed visualization and using a rigorous method, such as a pilot study, to help calculate the minimum sample size of the controlled study.

\section{Related Work}
\label{sec:related_work}
Our work is closely related to design study methodologies, and process models and guidelines thereof.

\subsection{Design Study and Its Applications}
Design study is an interdisciplinary research method in visualization.
It is mainly the design centered on domain expert users for specific domain problems. 
In interdisciplinary settings, design study tends to have a higher success rate than human-centered design (63\% vs. 25\%)~\cite{Wentzel2025}, and its core value lies in being problem-oriented, providing targeted solutions, and enhancing user experience through user-centered design, ensuring that visualization solutions better match user needs and expectations.

Design study methodology can be applied to a variety of domains, including medicine. 
The IRVINE visual analytics system was developed in close collaboration with automobile engineers to leverage interactive data labeling and clustering methods to facilitate the analysis of large amounts of acoustic data to detect and understand previously unknown errors in the electric engine manufacturing process~\cite{Eirich2022}.
RfX is a visual analysis system for analyzing the decision-making process of random forests. 
It allows expert automotive engineers without a data analysis background to identify the relationships in the feature subspace of random forests and detect hidden patterns in the underlying data of the model, which helps to change the electric engine test program and reduce the test time of components~\cite{Eirich2022a}.
Overview is an application for systematic analysis of large document collections based on document clustering, visualization and tagging, and this analysis goes beyond the journalism domain and involves the design and evaluation of other visualization tools~\cite{Brehmer2014}.
GEViTRec can handle a variety of dataset types and automatically generate visually coherent combinations of graphs, which can help genomic epidemiology experts identify and contain outbreaks of deadly diseases such as Ebola~\cite{Crisan2022}.
A web-based visualization tool named Trevo was developed for evolutionary biologists to analyze relationships between species, which can visually explore and analyze multivariate datasets and phylogenetic trees~\cite{Rogers2021}.

\subsection{Process Models and Practical Guidelines for Design Study}
\label{sec:models}

The four-level nested model by Munzner~\cite{Munzner2009} is a well-known model that can guide visualization design and validation.
Many later models related to visualization design have connections to this nested model.
Sedlmair et al. introduce a nine-stage framework -- learn, winnow, cast, discover, design, implement, deploy, reflect, and write~\cite{Sedlmair2012}. 
The framework is overall linear, indicating that one stage follows another, with some operations dependent on earlier stages.
The framework can provide some guidance for those conducting a design study and also serve as a starting point for further methodological discussions.

A typical design study can take months or even years to complete. 
Syeda et al. propose a design study ``Lite'' methodology (DSLM)~\cite{Syeda2020}, which is a simplified version of the nine-stage framework~\cite{Sedlmair2012}, following the nested model~\cite{Munzner2009} with additional preconditions and scope. 
The DSLM approach speeds up the design study process and makes it available to novice students within a 14-week time frame.

To fill the gap between the activities that visualization designers engage in and the visualization decisions they make, McKenna et al. propose the design activity framework, including four overlapping activities -- understand, ideate, make, and deploy~\cite{McKenna2014}. 
Unlike the nine-stage framework~\cite{Sedlmair2012}, this framework not only contains activities and methods, but also considers motivations, outcomes, and explicit connections to the nested model~\cite{Munzner2009}. 
By enabling and emphasizing workflows, the framework provides high flexibility, allowing it to capture the true nature of multilinear, real-world visualization design more completely than previous process models to help guide designers through the visualization design process.

Lam et al.~\cite{Lam2018} propose a framework designed to bridge the gap from high-level domain goals to specific low-level tasks, which helps locate analysis goals by placing each goal under the axes of specificity (explore, describe, explain, confirm) and number of data groups (single, multiple). 
With this framework, a visualization designer or researcher first determines the analysis goal for each unit of analysis in the analysis stream and then encodes the individual steps using existing task classifications.
Combining the target context, the level of specificity, and the number of groups participating in the analysis, it is possible to transform the questions asked and actions taken by the target users from the domain-specific language and context into a more abstract form, and then proceed to the next step of visualization design work.

In addition to the aforementioned models, some practical guidelines can guide researchers to conduct visualization and visual analysis design.
Shneiderman proposes that different visualization methods should be selected according to different tasks and data types in the process of information visualization~\cite{Shneiderman1996}. 
Roberts et al. propose the Five Design-Sheet (FDS)~\cite{Roberts2016} method, which enables users to iteratively create information visualization interfaces using a low-fidelity approach.
Kerzner et al. develop a framework for the Creative Visualization Opportunity (CVO) workshop that can help create outcomes that advance visualization methods in the early stages of visualization research~\cite{Kerzner2019}.
VizItCards is a card-driven workshop developed for information visualization that aims to provide practice in good design skills while reinforcing key concepts, to produce positive collaboration and high-quality design~\cite{He2017}.
Meyer et al. explore the nature of visualization design study from the perspective of interpretivism, and proposed a preliminary set of six criteria for rigor, aiming to guide researchers to construct, communicate and evaluate rigorous knowledge claims and get them to rethink how to conduct effective design study and learn new things in the process~\cite{Meyer2020}.

Although existing process models are available to guide design studies, they are generic as they aim at a wide range of application areas. 
To our knowledge, an operational process model that takes into account the characteristics of medical problems to guide a specific medical visualization design study does not exist yet. 
Therefore, this paper proposes a new process model.
A comparative analysis of our model and existing models can be found in~\cref{sec:model_comparison}.

\section{Design Study Process Model for Medical Visualization}
\label{sec:process_model_whole}

In this section, we first identify features of medical-related visualization and visual analysis works through a literature review. 
Subsequently, we introduce the design study process model for medical visualization based on the findings of the review.

\subsection{Features of Medical Visualization and Visual Analysis}
\label{sec:features}

\begin{figure}[htb]
    \centering
    \includegraphics[width=\linewidth, trim={0.2cm 0.1cm 0.4cm 0.1cm},clip]{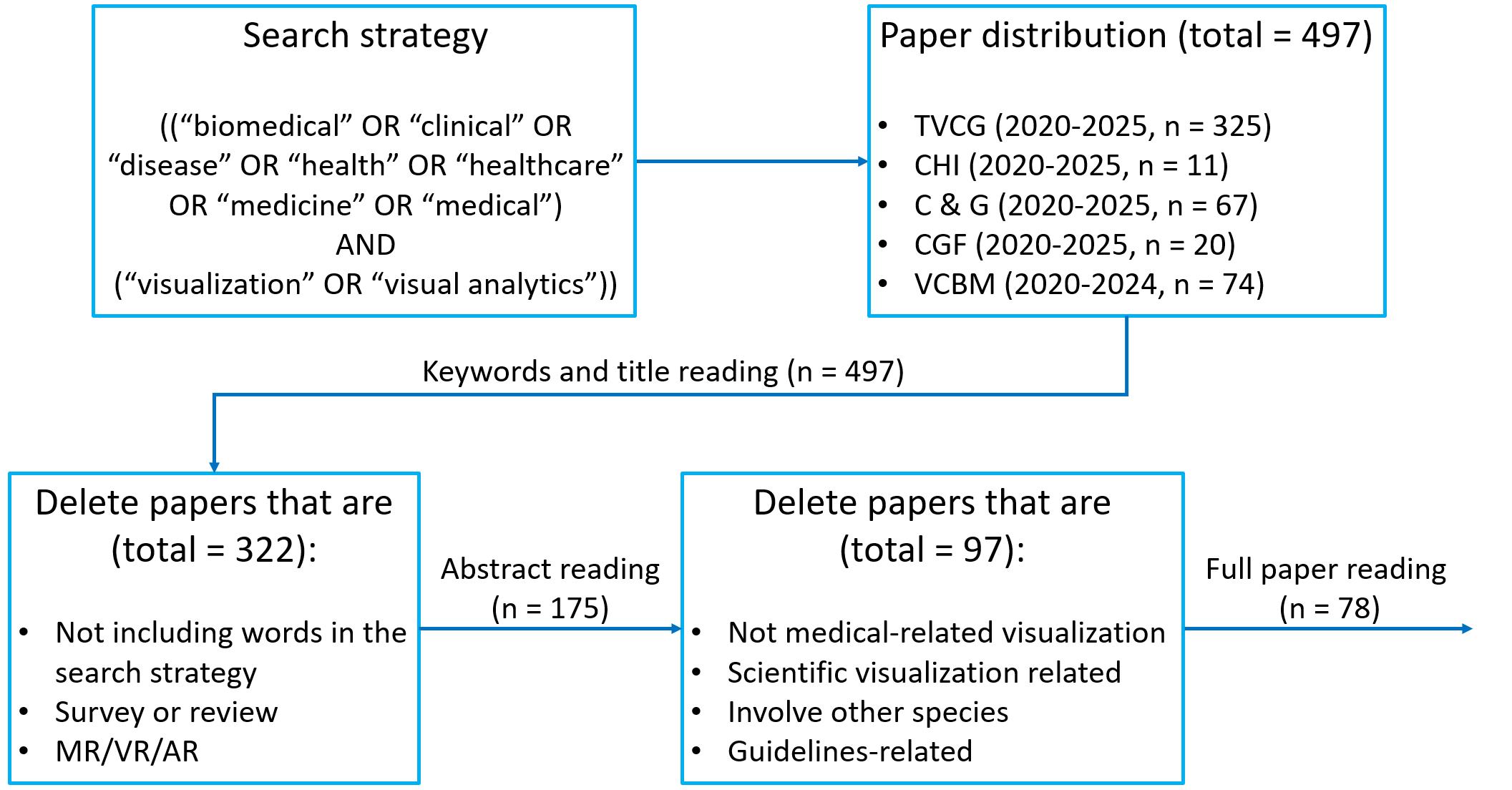}
    \caption{The workflow of the literature review process. Initial searches identified 497 articles (TVCG: 325, CHI: 11, C\&G: 67, CGF: 20, VCBM: 74). Title and keyword screening excluded 322 non-eligible articles. Abstract screening of the remaining 175 articles excluded a further 97, leaving 78 articles for full-text assessment.}
        % \vspace{-0.5em}
    \label{fig:literature_review_workflow}
\end{figure}

This study systematically reviewed literature on visualization and visual analytics addressing specific medical problems (``medical-related''). 
The two authors jointly discussed the inclusion and exclusion criteria, cross-validated the literature screening process, and collectively reviewed the included literature. 
When there were discrepancies in the coding of the literature, further discussions took place, leading to a final decision.
The workflow of the literature review process is shown in~\cref{fig:literature_review_workflow}.

We excluded review/survey papers, application papers related to mixed reality (MR), virtual reality (VR), and augmented reality (AR). 
Additional exclusion criteria were as follows: (1) Non-medical-related visualization; (2) Scientific visualization (such as focusing on image segmentation and calibration); (3) Studies involving non-human species; (4) Guidelines-related papers (e.g., discussing visualization roles or summarizing experimental insights).
We conducted a comprehensive literature search across five key publication venues. 
Using the search strategy~\cite{Ye2024a, Zhang2024} ((``biomedical'' OR ``clinical'' OR ``disease'' OR ``health'' OR ``healthcare'' OR ``medicine'' OR ``medical'') AND (``visualization'' OR ``visual analytics'')), we searched IEEE Transactions on Visualization and Computer Graphics (TVCG), the ACM CHI Conference on Human Factors in Computing Systems (CHI), Computers \& Graphics (C\&G), and Computer Graphics Forum (CGF), spanning the years 2020 to 2025. 
This search yielded 325, 11, 67, and 20 articles from these venues, respectively.
Additionally, we retrieved all papers published in the Eurographics Symposium on Visual Computing for Biology and Medicine (VCBM) between 2020 and 2024, obtaining 74 articles.
This initial search resulted in a combined total of 497 articles from these five sources.
Subsequent screening by titles and keywords refined this to 175 articles. 
Abstract review led to 78 articles retained for full-text assessment, with a primary focus on the methodological contributions detailed in each paper.

These works cover a wide range of data from molecular to individual to population levels, such as genes, proteins, computed tomography (CT), magnetic resonance imaging (MRI), electronic health records, patient self-recorded data, prescription data, population mobility data~\cite{Somarakis2021a, Warchol2023, Meuschke2023, Jessup2022, Corvo2021, Furmanova2021, Pandey2020, Morth2025, Ouyang2024, Opaleny2025, Stahlbom2025, Choi2024, Joensson2020, Sondag2022, Hoehn2024, Raj2023, El-Sherbiny2023}, etc.
They are designed mainly for insight analysis and decision support~\cite{Somarakis2021a, Choi2024, Schaefer2024, Pandey2023, Wentzel2025, Yan2025, Kim2025, Scimone2024, Jiang2024, Zhou2023a, Wang2023a, Joensson2020, Sondag2022, Eulzer2021a, El-Sherbiny2023, Pereira2023} or medical education and information communication~\cite{So2021, Meuschke2022, Ouyang2024, Raj2023, Jiaan2025, Kleinau2022}.
People involved in these papers are mainly medical professionals who have background knowledge of a specific medical problem, for example, surgeons, radiologists, (bio)medical experts, microscopists, pathologists, 
% oncologists, physicians, biologists, structural biologists, computational scientists, protein engineers, genomics scientists, neuroscientists, medicinal chemists, pediatric oncology clinicians, nurses, novice physicians and interns, neurologists, health insurance audit experts, policymakers, 
etc.~\cite{Warchol2023, Meuschke2023, Jessup2022, Morth2025, Jiang2024, Zhou2023a, Wang2023a, Joensson2020, Sondag2022, Hoehn2024, Raj2023, Eulzer2021a}.
Some of them also involve non-medical professionals such as patients, the general public~\cite{So2021, Meuschke2022, Scimone2024, Raj2023, Jiaan2025}, and semi-professionals such as medical students~\cite{Ouyang2024, Xu2025}.

Through detailed reading and analysis of these papers, we identify three features that have not been discussed or represented in existing process models.

\begin{itemize}
    \item \textbf{A medical problem may involve multiple stakeholders.}
            For example, when using multiplexed tissue image data to describe tumor characteristics, it is necessary to consider the different needs of cell biology experts and pathologists~\cite{Jessup2022}.
            Narrative-driven visualization that involves both clinicians and the elderly in iterative design can help the elderly understand medical information~\cite{Jiaan2025}.         

    \item \textbf{The analysis and resolution of certain medical problems may need to be carried out in stages.}
    For example, the protein loop grafting process is divided into six stages according to the actual workflow to assist protein engineers in insight analysis~\cite{Opaleny2025}.
    Analyze the data of patients with acute ischemic stroke according to different stages of the analysis pipeline to drive clinical decision-making~\cite {Kim2025}.

     \item \textbf{Some medical problems focus on differences between subgroups.}
    For example, explore the differences between healthy volunteers and patients with a pathologically altered aorta~\cite{Meuschke2023}, compare the differences in brain activity among different patient groups~\cite{Joensson2020}.
\end{itemize}

\begin{table}[htbp]
  \centering
  \caption{Design features of the reviewed medical visualization works.}
  \label{tab:medical_features}
  \begin{tabular}{lr}
    \toprule
    \textbf{Features} & \textbf{Number of papers} \\
    \midrule
    Multiple stakeholders &  62 \\
    Multiple stages &   33 \\   
    Multiple subgroups &   38\\ 
    \bottomrule    
  \end{tabular}
\end{table}
In~\cref{tab:medical_features}, we summarize the frequencies of these features of the 78 reviewed papers.

\subsection{Process Model for Medical Visualization}
\label{sec:process_model}
Considering the characteristics of medical visualization and visual analysis work (\cref{sec:features}), along with our own research experience and insights from different design study process models and practical guidelines described in previous work (\cref{sec:models}), we propose a design study process model suitable for medical visualization as shown in~\cref{fig:process_model}.
The steps of this model are as follows: (1) select collaborators and identify domain problems; (2) identify stakeholders; (3) locate the analysis goals; (4) dismantle the analysis goals; (5) design visualization; (6) implement prototype; (7) evaluate; (8) promote.
In the remainder of this section, we explain each step of the model in detail.

\subsubsection{Select Collaborators and Identify Domain Problems}
\label{sec:domain_problem}
Design study is typically problem-driven visualization research conducted in collaboration with domain experts~\cite{Sedlmair2012}.
In our experience, we have communicated with many medical experts about potential collaborations during our daily work. However, most of these interactions tend to be superficial and rarely lead to valuable collaborations.
Collaborators should be selected before the design study officially begins. 
The criteria for selection should include: these medical experts focus on addressing specific medical issues, can obtain appropriate data, and believe in the potential of visualization.
Once the collaborators are identified, further discussions are needed to identify the specific medical problem to address.

Semi-structured interviews~\cite{Xu2025} are a commonly used way to figure out the real thoughts and demands of the collaborator, which can help identify domain problems.

\subsubsection{Identify Stakeholders}
\label{sec:stakeholders}

Medical problems may involve multiple stakeholders (e.g., role 1, role 2, $\ldots$ role $m$), such as doctors, nurses, and patients. 
Stakeholders in this paper refer to anyone concerned with a specific medical problem, including the collaborators selected in~\cref{sec:domain_problem}, other than the visualization researchers, who are an integral part of any design study project.
Different stakeholders have various expertise, backgrounds, needs, and perspectives, and may have different opinions on problem definition and solution~\cite{Rajabiyazdi2021}, all of which pose challenges to visualization design~\cite{Mahyar2020, Burns2023}.
Researchers have recognized the importance of considering individual and group differences in visualization systems, rather than relying on a ``one-size-fits-all'' interface~\cite{Liu2020}. 

Before conducting a specific medical visualization design study, it is necessary to repeatedly and deeply discuss with collaborators to jointly identify the different stakeholders involved in a specific medical problem.
Oppermann et al. find that some design studies are initiated primarily by obtaining interesting real-world datasets rather than selecting specific stakeholders, and propose the concept of data-first design studies~\cite{Oppermann2020}. 
In data-first design studies, early selection of data can limit the appropriate selection of stakeholders, who are selected based on whether the selected data source can support their tasks.
Stakeholder identification can also be data-driven, that is, finding potential stakeholders based on available data.

Further analyzing the 62 works involving multiple stakeholders, we find that the relevant tools or systems are designed for a specific role, that is, focused on the target user, and not always for all stakeholders, especially when involving multiple stakeholders with inconsistent levels of expertise.
Different levels of medical expertise of stakeholders and target users (professionals, semi-professionals, and non-professionals) can correspond to the above-mentioned three main medical-related visualization applications (\cref{tab:stakeholders_target}).

\begin{table}[htbp]
  \centering
  \caption{Stakeholders vs. Target Users}
  \label{tab:stakeholders_target}
  \small
  \begin{tabularx}{0.48\textwidth}{
    >{\hsize=1.0\hsize\RaggedRight}X  % 20%宽度，左对齐
    >{\hsize=0.8\hsize\RaggedRight}X  % 50%宽度，左对齐（适合长文本）
    >{\hsize=1.2\hsize\RaggedRight}X   % 30%宽度，居中对齐
  }
    \toprule
    \textbf{Stakeholders} & \textbf{Target users} & \textbf{Applications}\\
    \midrule
    various backgrounds professionals & professionals as a whole & insight analysis and decision support \\
    \\
    professionals and semi-professionals & semi-professionals (e.g., students) & medical education \\
     \\
    professionals and non-professionals & non-professionals (e.g., patients) &  information communication \\
    \bottomrule
  \end{tabularx}
\end{table}

Implicit in selecting stakeholders is the need to specify the target users.
For instance, a patient-doctor communication requirement for a clinical procedure is raised by the collaborating specialist doctor (collaborator).
Here, different stakeholders may be involved, such as doctors of varying levels of expertise and the patients. 
However, the target users of the tools designed for this case can be either doctors (to assist in managing the entire disease process) or patients (to help them understand the occurrence and development of the disease, etc.).

\subsubsection{Locate the Analysis Goals}
\label{sec:analysis_goals}

After the stakeholders are identified, it is necessary to clarify what each stakeholder considers to be the objectives of the analysis for a particular problem.
For example, role 1 has goals 1,2,3, role 2 has goals 2,4,5,6, etc.  (\cref{tab:analysis_goals}).
Combining the goals of different roles yields the final analysis goals for the domain problem identified in~\cref{sec:domain_problem}.

\begin{table}[htbp]
  \centering
  \caption{An example of analysis goals for different stakeholders}
  \label{tab:analysis_goals}
  \small
  \begin{tabular}{ll}
    \toprule
    \textbf{Stakeholders} & \textbf{Analysis Goals} \\
    \midrule
    role 1 & goals 1,2,3 \\
    role 2 & goals 2,4,5,6 \\
    $\ldots$ & $\ldots$ \\
    role $m$ & goals 3,5,$\ldots$, $n$ \\
    \midrule
     roles 1,2,3,$\ldots$, $m$ & goals 1,2,3,$\ldots$, $n$ \\
    \bottomrule
  \end{tabular}
\end{table}

Organizing different stakeholders through workshops~\cite{Kerzner2019} to facilitate discussions, or conducting semi-structured interviews~\cite{Xu2025} to interview each individual, are viable methods for collecting stakeholders' objectives.

\subsubsection{Dismantle the Analysis Goals}
\label{sec:requirements}

Analysis goals can be divided into different stages depending on the target users.
As shown in~\cref{tab:stakeholders_target}, applications for professionals are mainly \emph{insight analysis and decision support}.
The solution of a specific medical problem may need to be carried out gradually in stages according to the \emph{analytic logic}, and the analysis of the subsequent stage needs to depend on the analysis of the previous stage. 
The applications for semi-professionals, such as students, are \emph{medical education}, and two levels can be applied to distinguish the different stages.
Students can acquire knowledge passively, which may require the gradual design of visualizations in the order of \emph{cognitive habits}, such as from easy to difficult, from diagnosis to treatment to prognosis, etc.
Students can also be active in acquiring knowledge, which involves a data analysis process, so that the visualization can be designed stage by stage according to \emph{analytic logic}.
Applications for non-professionals, such as patients or the general public, are \emph{information communication}.
Designing and conveying knowledge and information in line with the \emph{cognitive habits} is beneficial for visualization design.
The works reviewed are also divided into stages at these two levels.

\begin{table}[htbp]
  \centering
  \caption{Dismantle analysis goals into stages and tasks, considering target users and subgroups. }
  \label{tab:dismantle_goals}
  \small
  \begin{tabularx}{0.48\textwidth}{
    >{\hsize=1.0\hsize\RaggedRight}X  % 25%宽度，左对齐
    >{\hsize=1.0\hsize\RaggedRight}X  % 30%宽度，左对齐（适合长文本）
    >{\hsize=0.5\hsize\RaggedRight}X   % 25%宽度，居中对齐
    >{\hsize=1.5\hsize\RaggedRight}X   % 20%宽度，居中对齐
  }
    \toprule
    \textbf{Target users} & \textbf{Stages} & \textbf{Subgroups} & \textbf{Tasks type}\\
    \midrule
    professionals & analytic logic & yes & inferential: hypothesis-based (recommended) \\
     &  & no & inferential  \\
    semi-professionals & active learning: analytic logic & yes & inferential: hypothesis-based  (recommended) \\
     & active learning: analytic logic & no & inferential  \\
     & passive learning: cognitive habits & no & descriptive \\
    non-professionals & cognitive habits & no &  descriptive \\
    \bottomrule
  \end{tabularx}
\end{table}

Previous process models~\cite{Munzner2009, Sedlmair2012, Lam2018} propose to abstract domain problems into tasks.
In our model, we believe it is necessary to further distinguish different analysis tasks within each stage.
Specifically, when focusing on passive learning by students or information communication by non-professionals, different stages can be distinguished according to cognitive habits, at which point we define the task type within the stage as \emph{descriptive}.
When targeting active learning by students or insight analysis and decision support by professionals, different stages can be distinguished according to the analytic logic, at which point we define the task type within the stage as \emph{inferential}.

In medicine, a specialized category of inferential tasks involves the analysis of differences among multiple subgroups, such as healthy versus pathological cohorts. 
Hypothesis-driven tasks encompass the formulation of testable hypotheses regarding specific conditions, followed by the design of experiments to either validate or refute these hypotheses. 
When a task involves multiple subgroups, using a hypothesis-driven approach and designing visualizations based on possible differences among different groups can help develop effective methods to compare these differences and drive decision support. 
We therefore recommend that inferential tasks can be further distinguished as \emph{hypothesis-based} or \emph{hypothesis-free}, depending on whether different subgroups are involved.
For those defined as \emph{inferential}, if they involve different subgroups, their analysis may benefit from a \emph{hypothesis testing}.
We define this type of analysis as \emph{hypothesis-based}; Otherwise, \emph{hypothesis-free}.

This decomposition of analysis goals into stages according to different target users and further differentiation of different tasks depending on whether subgroups are involved could simplify the analysis and solution of medical problems and speed up the visualization design process.

\subsubsection{Design Visualization}
\label{sec:visualization_techniques}
Intuitive and effective visualizations help users learn quickly~\cite{Zhou2023a}.
Factors such as data type, purpose and audience, data volume and granularity, complexity of data, relationships and patterns of data, spatial properties of data, clarity and conciseness, and required interactivity need to be considered when choosing the appropriate data visualization technique to make informed decisions~\cite{Heer2010, Shneiderman1996, Few2009}.

Many popular data visualization techniques applicable to various domains are used in medical visualization-related works. 
For example, bar charts to show specific values for each category~\cite{Warchol2023, Corvo2021, Pandey2023}, histograms to show the frequency of a particular event~\cite{Meuschke2023, Jessup2022}, line charts to track the evolution of variables over time~\cite{Morth2025, Pandey2023}, heat maps to display the density or intensity of data points~\cite{Warchol2023, Meuschke2023, Morth2025}, scatter plots to demonstrate the relationship between two continuous variables~\cite{Warchol2023, Corvo2021, Stahlbom2025}, box plots to reflect the distribution of variables~\cite{Warchol2023, Corvo2021}, violin plots also used to reflect the distribution~\cite{Meuschke2023}, network charts to represents the connection between data points\cite{Jiang2024}, parallel coordinates to visualize multivariate data~\cite{Meuschke2023, Corvo2021}, and interactive techniques to explore patterns in data~\cite{Meuschke2022, Warchol2023, Meuschke2023, Jessup2022, Pandey2020, Morth2025, Opaleny2025, Schaefer2024}.

In the process of medical visualization design studies, the practical guidelines mentioned in~\cref{sec:models}, such as FDS~\cite{Roberts2016} and VizItCards~\cite{He2017}, can be combined to select different visualization techniques or combine multiple techniques~\cite{Warchol2023, Meuschke2023, Jessup2022, Pandey2020, Morth2025, Opaleny2025, Brandt2025} to help create expressive visualizations according to different medical problems, target users, analysis tasks and data types. 
There is also a need to design and evaluate new visual encodings and visualization techniques~\cite{Meuschke2022, Fan2025, Morth2025, Stahlbom2025, So2021} when necessary.
In addition, it is sometimes necessary to devise algorithms to improve computation and rendering efficiency~\cite{Jessup2022, Choi2024, Pandey2023}.
Well-designed algorithms play a key role in data preprocessing and optimization, visualization generation, and interaction enhancement~\cite{Maaten2008, Heer2012, Ye2024}, which can improve the efficiency and depth of data analysis, help users extract valuable information from complex data, and improve decision support.
Combining visualization techniques, visual encodings, interaction techniques, and efficient algorithms can help achieve effective visualization design for specific medical problems.

Visualization design is a subjective process that emphasizes the importance of iterative design, which is in line with the design recommendations made by Opaleny et al., who argue that the iterative design process is important to find a representation that conforms to the conventions and requirements of the application domain~\cite{Opaleny2025}.
Visualization designers can iterate the design based on feedback from the evaluation step (\cref{sec:evaluation}), focusing on whether different visualization designs meet the analysis requirements of different tasks.

\subsubsection{Implement Prototype}
\label{sec:prototype}
Aforementioned visualization techniques or algorithms for different purposes do not exist in isolation. 
They need to be integrated to form an interactive visualization interface, i.e., a visualization prototype. 

While visualization can reveal data patterns, overly complex interfaces may increase cognitive load and hinder the usability of the system, especially in scenarios involving interaction across multiple rounds of iterations during toolkit design~\cite{Wang2024}.
Therefore, in the design of specific visualization interfaces, it is possible to consider providing a multi-level view to visualize complex high-dimensional data and support progressive analysis.
For example, FraudAuditor~\cite{Zhou2023a} divides the analysis task and coordination view into three levels -- overview level, group level, and patient level.
Researchers may also consider creating multiple cascaded views based on the logical order of the analysis process or cognitive habits (\cref{sec:requirements}). 
In addition, different prototypes can be developed for various stakeholders (\cref{sec:stakeholders}).

In specific medical visualization design studies, analysts can also refer to the FDS~\cite{Roberts2016} and VizItCards~\cite{He2017} methods to first design prototype sketches, then create the prototype, and iterate the prototype when necessary based on the evaluation results (\cref{sec:evaluation}). 
The iteration can focus on whether the functions designed for the same stage or for a specific target user population are well integrated into the prototype.
The ideal prototype should reasonably integrate all the task analysis requirements of the specific target users, and distinguish views by stages, presenting them in the form of a single interactive interface or multiple interfaces.

\subsubsection{Evaluate}
\label{sec:evaluation}

Value-driven visual evaluation~\cite{Stasko2014} argues that the evaluation of a technology or system should be able to identify and illustrate its value, and should also be able to potentially improve the technology and system being built. 
After completing the visualization design and prototype design under the various visualization design requirements formed in the goal dismantling (\cref{sec:requirements}) step, it is necessary to use appropriate methods to evaluate the effectiveness of visualization according to different target users, analysis goals, problems, etc.
Researchers can also continue to iterate and improve with user feedback to obtain a satisfactory visualization solution. 
This iterative design process has the potential to alternate between visualization design (including prototype design) and evaluation, which is reflected in~\cref{fig:process_model} as bidirectional and reverse arrows.

Various types of methods are available to evaluate visualization designs.
Dashboard Comparison effectively evaluates visualization efficacy by analyzing internal structure, functionality, accuracy, and interactivity when parameters of different dashboards are closely aligned~\cite{Islam2024}.
Insight-based Evaluation measures visualization effectiveness by capturing user insights through verbal reports or task-based feedback, enabling design comparison and refinement~\cite{North2006}.
Heuristic Evaluation is a method of analyzing usability where evaluators assess an interface to identify specific design issues impacting user experience~\cite{Nielsen1990}.
Eye-tracking evaluates usability by analyzing gaze patterns, attention distribution, and visual search strategies to optimize interface design and information prioritization~\cite{Blascheck2016}.
Field Observation studies real-world user interactions to derive design requirements and improve visualization tools based on observed workflows~\cite{Rae2022}.
Interviews and Focus Groups employ think-aloud techniques to gather participant feedback on the evaluation of visualization comprehension, perception, and design~\cite{Lewis2006}.
Standardized questionnaires (e.g., SUS~\cite{Brooke1996}, NASA-TLX~\cite{Hart1988}, Likert~\cite{Likert1932} scales) collect user feedback on visualization satisfaction and usability, while statistical metrics (e.g., task completion time, accuracy) provide quantitative evaluation for design optimization.
Algorithmic performance in visualization is evaluated through numerical metrics such as runtime, false positive rate, false negative rate, and scalability.

\begin{table}[htbp]
  \centering
  \caption{Evaluation methods commonly used in medical visualization.}
  \label{tab:evaluation_methods}
  \small
  \begin{tabularx}{0.48\textwidth}{
    >{\hsize=0.6\hsize\RaggedRight}X  % 20%宽度，左对齐
    >{\hsize=1.5\hsize\RaggedRight}X  % 50%宽度，左对齐（适合长文本）
    >{\hsize=0.9\hsize\RaggedRight}X   % 30%宽度，居中对齐
  }
    \toprule
    \textbf{Evaluation methods} & \textbf{Possible combined methods} & \textbf{Evidence level in medicine} \\
    \midrule
    Controlled studies & Various methods can be combined, such as Dashboard Comparison, Insight-based Evaluation, Heuristic Evaluation, Eye-tracking, Interviews and Focus Groups, Standardized Questionnaires & \textbf{Highly recognized} (quantitative or qualitative, and usually quantitative) \\
    \\
    Case studies & Multiple methods can be combined, such as Insight-based Evaluation, Interviews, and Focus Groups & Less common (usually qualitative) \\
     \\
    User \newline evaluations & Multiple methods can be combined, such as Dashboard Comparison, Insight-based Evaluation, Heuristic Evaluation, Interviews and Focus Groups & Less common (usually qualitative) \\
    \bottomrule
  \end{tabularx}
\end{table}

In most cases, a single evaluation method is not sufficient for the evaluation of visualization design, and multiple evaluation methods need to be combined~\cite{Jessup2022, Pandey2023, Jiang2024}.
Commonly used comprehensive evaluation methods (see \cref{tab:evaluation_methods}) are controlled studies~\cite{Akpan2019}, case studies~\cite{Shneiderman2006}, and user evaluation~\cite{Greenberg2008}, which combine and take advantage of various evaluation methods, not limited to the ones mentioned above.
Controlled studies, which can be a comparison between two groups or a comparison between the designed system and an existing system, are highly recognized evaluation methods in the medical domain.

In the process of a specific medical visualization design study, researchers should examine the reliability, validity, and sensitivity of evaluation indicators in capturing desired results and accurately assessing the performance of visual analytics systems~\cite{Islam2024}.
Reflecting on evaluation goals and problems before choosing a specific evaluation method can provide a new space for people to evaluate visualization effectively and efficiently. 
According to the specific domain problem or goal, one or more evaluation methods can be applied, and the results obtained by analyzing and synthesizing various evaluation strategies can illustrate the value of visualization or enhance the current visualization system development.

\subsubsection{Promote}
\label{sec:promotion}

A visualization design that can be promoted has great value of universality, which can reduce future design efforts. 
In medicine, visualization design targeted at one problem can be promoted to populations sharing certain characteristics or other problems involving similar tasks.
For example, in applications of medical communication, visualizations that are found to be effective for target users can be promoted to other populations with similar characteristics, such as whether they have a specific disease. 
In applications for insight analysis and decision support, effective designs for specific analysis tasks, such as survival analysis, can be promoted to other medical problems involving the task.

By summarizing experience and reflecting on the lessons learned, visualization design guidelines can be developed and can then be generalized to more similar visualizations.
The promotion process is optional and depends on the application positioning of the design study.

\subsection{Summary of the Model}
\label{sec:summarization}
The new design study process model proposed in this paper aims to provide a theoretical framework and practical guidance for visualization and visual analysis work related to medicine.
It can help researchers with issues such as how to identify different stakeholders, how to analyze problems and dismantle objectives, how to choose and design the appropriate visualization techniques, and how to choose or combine different evaluation methods.

\section{Use Cases}
\label{sec:examples}
In this section, we introduce the application of the proposed model to guide one of our previous works and reanalyze three medical visualization works using this model.
Figures in this section in larger sizes can be found in the supplemental material.

\subsection{Guide: Multi-outcome Causal Graphs Analysis}
\label{sec:multi_outcome}

\begin{figure}[htb]
    \centering
    \includegraphics[width=0.8\linewidth, trim={0.1cm 0.1cm 0.11cm 0.1cm},clip]{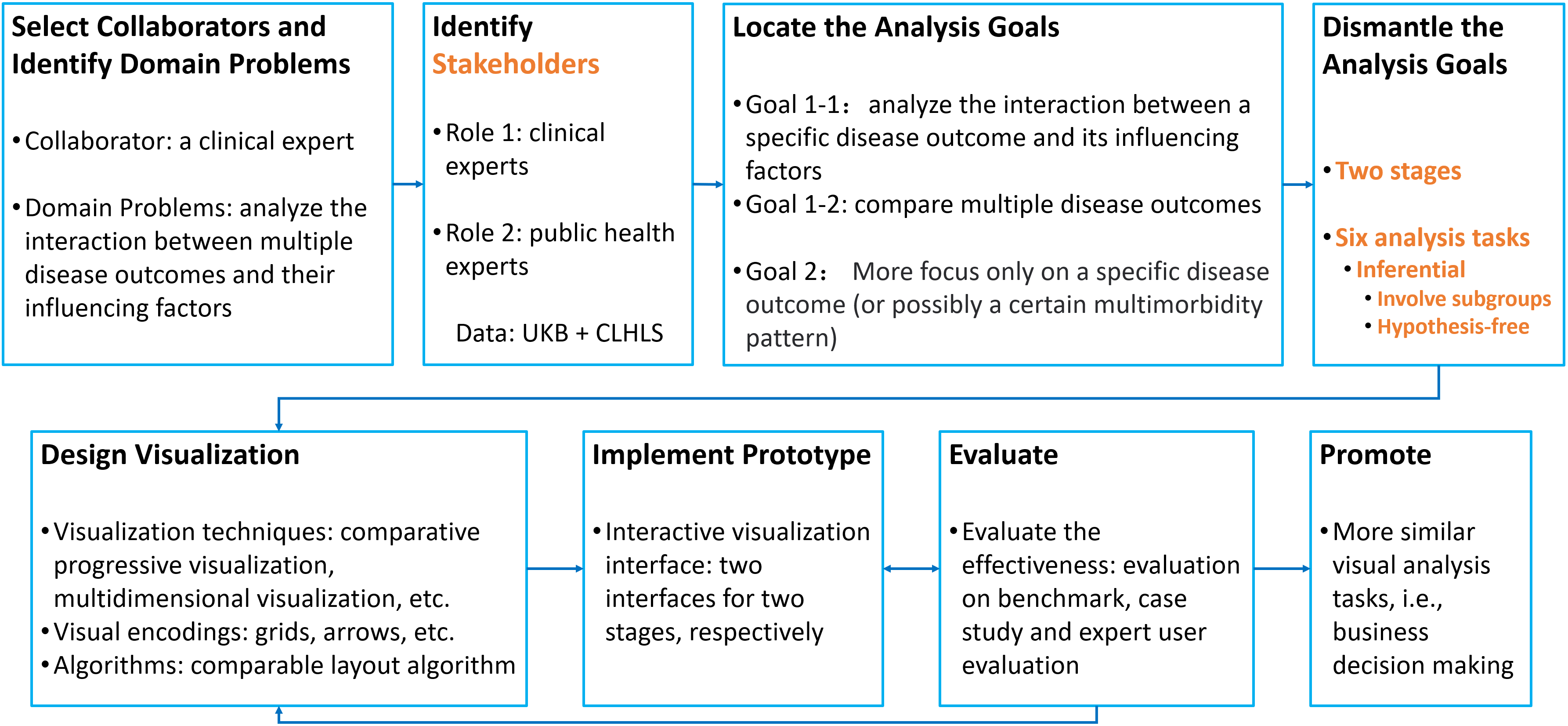}

    \caption{The workflow of \emph{visual analysis of multi-outcome causal graphs} guided by the proposed design study process model.}
        % \vspace{-0.5em}
    \label{fig:multioutcome_process}
\end{figure}

We applied this model to guide the visual analysis of multi-outcome causal graphs~\cite{Fan2025}.
In collaboration with a clinical expert, we first define the domain problem as analyzing interactions between multiple disease outcomes and their influencing factors. 
Based on this problem, we select two relevant datasets, identify the stakeholders facing this problem (clinical and public health experts), and determine their respective analytical objectives.
Further refinement through stakeholder engagement decomposes these goals into two distinct analysis stages and six specific tasks, forming the system's design requirements. 
To address these requirements, we design and implement targeted visualization techniques (e.g., comparative, progressive, multidimensional visualization), appropriate visual encodings (e.g., grids, arrows), a novel comparable layout algorithm, and dedicated interactive interfaces for each stage.
The effectiveness of the system is rigorously evaluated using quantitative metrics on benchmark data, a qualitative case study ($N$ = 1), and expert user evaluation ($N$ = 3).
The whole process of the realization of this work is illustrated in~\cref{fig:multioutcome_process}.

\begin{figure}[htb]
    \centering
    \includegraphics[width=0.8\linewidth, trim={0.1cm 0.1cm 0.11cm 0.1cm},clip]{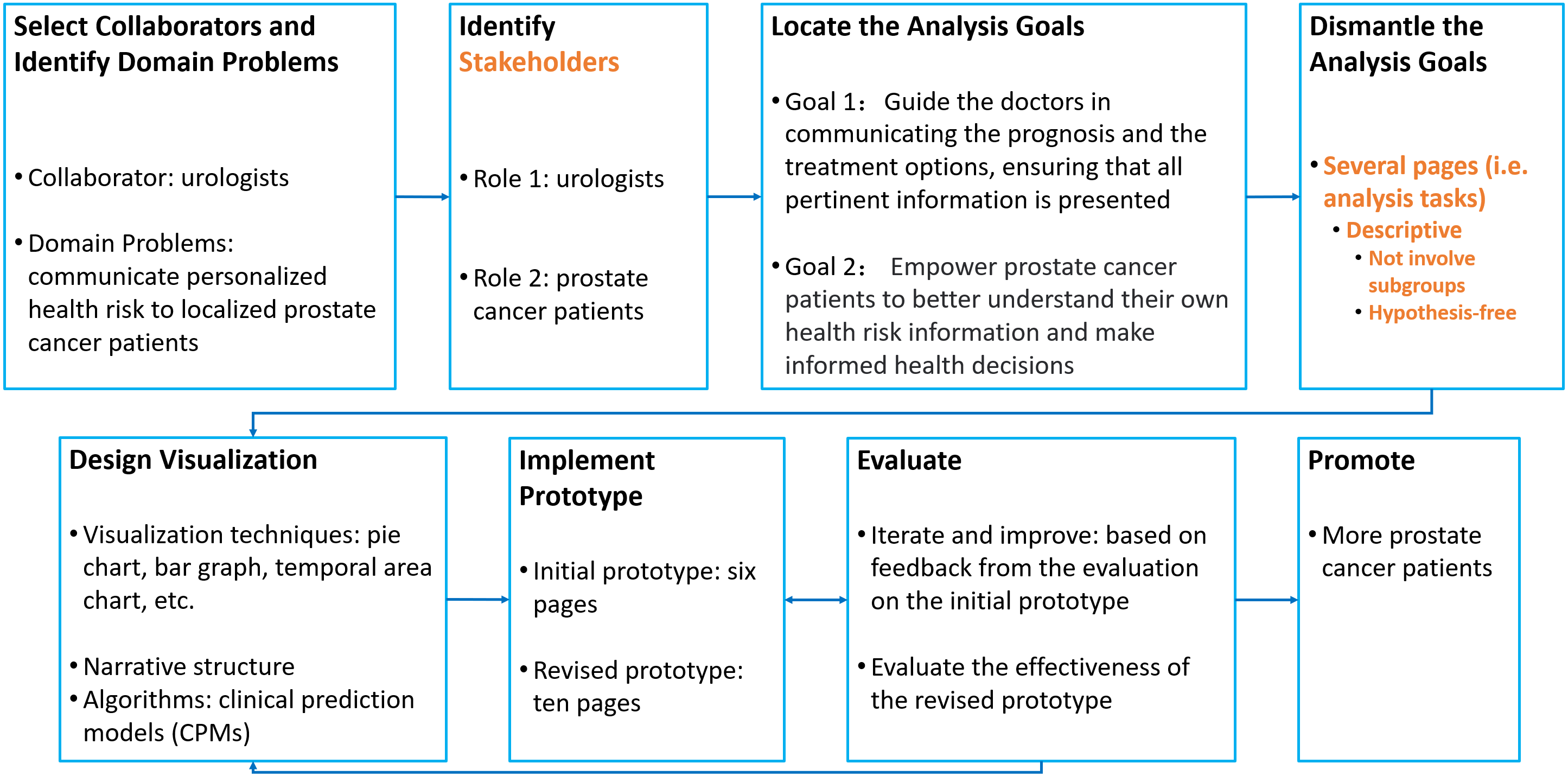}

    \caption{The workflow of the iterative design of PROACT reanalyzed under the proposed design study process model.}
        % \vspace{-0.5em}
    \label{fig:proact_process}
\end{figure}

\rev{
This study aims to explore the interactions among various specific diseases and their influencing factors.
From this perspective, the specific disease combinations themselves are regarded as a category of ``health outcomes'' worthy of attention. 
This requires that the visualization analysis method we develop should simultaneously meet the needs of public health experts (who focus on group patterns and prevention) and clinical experts (who focus on individual diagnosis and comparison).
The key premise of design is to recognize the differences and complementarities in the perspectives of various stakeholders. 
If the visualization solution only focuses on public health experts, it may weaken the detailed comparison function among diseases, which is the function that clinical decision-making frequently relies on; conversely, if it overly leans towards the perspective of clinical experts, it may neglect the overall pattern insight of multimorbidity, which is the focus of public health research.
Therefore, at the initial stage of design, it is necessary to proactively identify and integrate the needs of these two types of users, which is the foundation for ensuring that the visualization solution is both comprehensive and targeted.
}

\rev{
To achieve a systematic comparison of the complex relationships between various disease outcomes and their influencing factors, it is necessary to first accurately understand the individual outcome-factor relationships, indicating the existence of sequential dependencies. 
We followed the proposed phased progressive design model and divided the visualization process into two stages based on internal logic, ensuring that the second stage is built on the solid foundation of the previous one. 
In response to the demand for comparisons emphasized by clinical experts (such as the characteristics, treatment paths, and outcome comparisons of different diseases), under the guidance of this model, we have developed visualization techniques and interactive algorithms that support comparisons and causal analysis, aiming to present complex comparison issues intuitively and help users efficiently discover differences and patterns.
In conclusion, by constructing a visual analysis framework that can reconcile the perspectives of the public health community and individual clinical cases, this work provides effective support for interdisciplinary collaborative research.
}

\subsection{Reanalyze: PROACT}
\label{sec:proact}

We reanalyze the design of PROACT~\cite{Hakone2017}, a visualization tool for effectively communicating health risks to prostate cancer patients, using our model (\cref{fig:proact_process}).
There, researchers first selected urologists as their collaborators and determined the domain problem of communicating personalized health risks to localized prostate cancer patients.
Urologists and prostate cancer patients are naturally determined as stakeholders facing this problem.
Through two rounds of discussions with these stakeholders, mainly urologists, researchers determined two analysis goals and designed an initial six-page prototype, which combines visualization techniques (e.g., pie charts and bar charts), a clear narrative structure, and clinical prediction models to support risk computation. 
Pages here can correspond to analysis tasks in our model.
It can also be treated as stages, due to the requirement of a reasonable narrative structure.
Then they recruited both patients ($N$ = 6) and doctors ($N$ = 2) to evaluate the initial prototype via a semi-structured interview.
Feedback led to an iterative redesign, resulting in a revised ten-page prototype. 
A second evaluation with 6 new patients and the same 2 urologists assessed its effectiveness.

\rev{
In the original design of the PROACT health risk communication tool, although it ultimately involved both doctors and patients as the two types of stakeholders, its initial design stage primarily focused on the perspective and needs of doctors. 
The patients' feedback was not taken into consideration until the evaluation stage and was used to guide the iterative design of the subsequent visualization. 
Practice has shown that if, at the design stage, we can simultaneously and fully consider the differentiated needs of these two core groups using the model we proposed, it is expected to optimize the design at the source, effectively reducing subsequent revisions due to misalignment of needs or cognitive biases, and thus saving development resources.
}

\rev{
Although this tool is related to both doctors and patients, its core target users are the patient group.
Therefore, when designing the solution, the researchers ultimately followed the cognitive logic and information processing habits of the patients, adopted a clear narrative structure and simple visualization design to organize and present the content. 
This design strategy, which aligns with our model's view that categorizes stages based on cognitive habits, significantly enhances the comprehensibility of the visual content and the clarity of the interface, thereby ensuring effective communication.
}

\subsection{Reanalyze: e-MedLearn}
\label{sec:medlearn}

\begin{figure}[htb]
    \centering
    \includegraphics[width=0.8\linewidth, trim={0.1cm 0.1cm 0.11cm 0.1cm},clip]{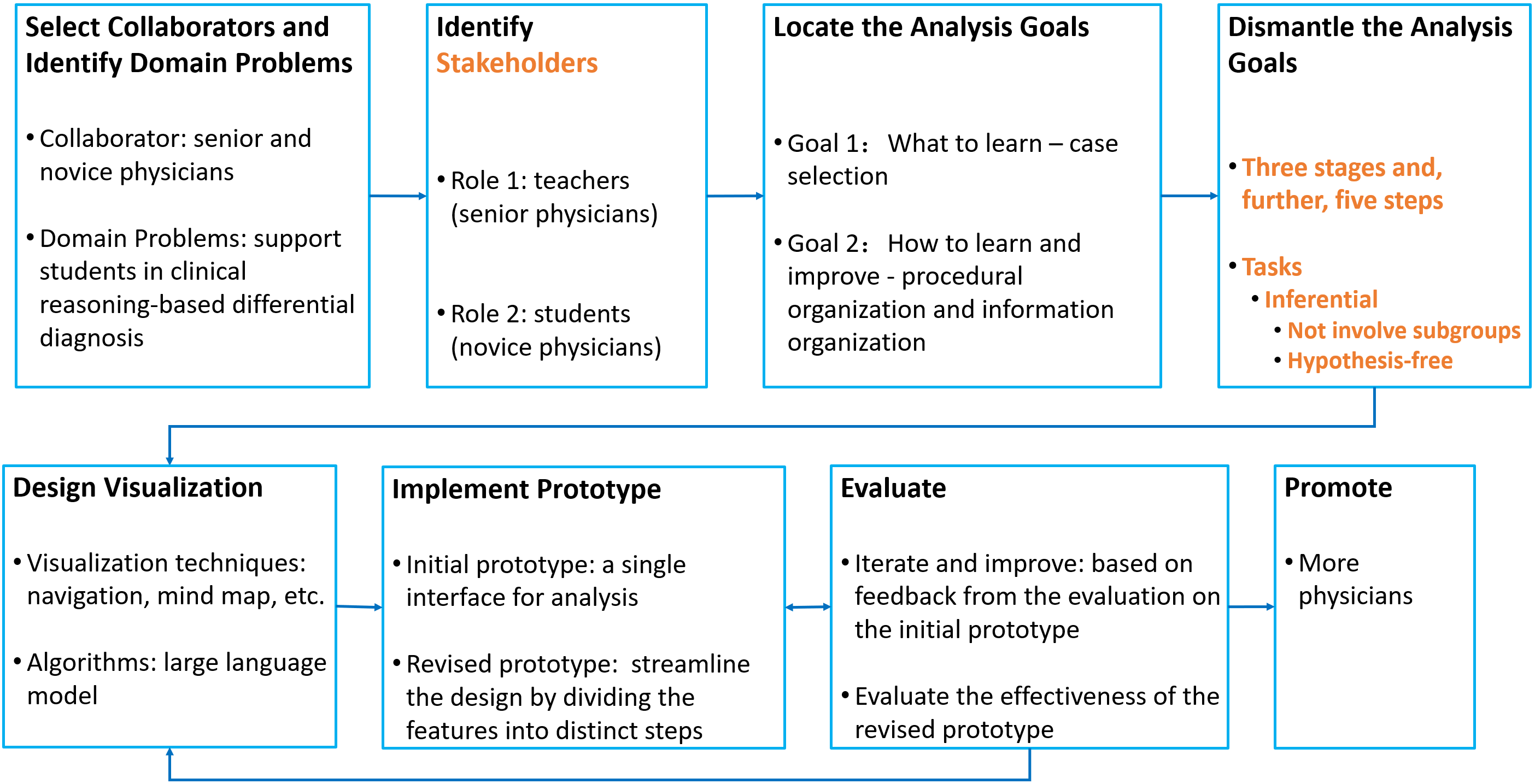}

    \caption{The workflow of e-MedLearn reanalyzed under the proposed design study process model.}
        % \vspace{-0.5em}
    \label{fig:emedlearn_process}
\end{figure}

The design of e-MedLearn~\cite{Xu2025} is reanalyzed as the third example (\cref{fig:emedlearn_process}).
e-MedLearn is a tool for medical education targeted at novice physicians.
Researchers collaborated with both senior (teachers) and novice (students) physicians to define the domain problem as supporting students in problem-based learning (PBL), with a focus on clinical reasoning-based differential diagnosis.
Through stakeholder discussions, they identified needs and barriers, shaping analysis goals. 
These goals informed a three-stage, five-step process structure. The solution features diverse visualizations (navigation, mind map, etc.) integrated with a large language model to support different functions. 
Design iterated from an initial single-interface prototype to a revised version with step-specific features. Effectiveness was evaluated through a two-stage study: a controlled study ($N$ = 18, 9 e-MedLearn vs. 9 baseline) and a testing interview ($N$ = 13, 10 novice and 3 senior).

Taking into account the needs and concerns of both senior and novice doctors, the designed tool has a wide applicability. 
Imagine this scenario: if only the needs of novice doctors were considered, the visualization design might only focus on how to learn and improve, while neglecting the need to first select appropriate cases for learning based on one's ability level. 
Moreover, designing different functions step by step according to analytical stages is conducive to a logical and reasonable visual analysis process.

\subsection{Reanalyze: GUCCI}
\label{sec:GUCCI}

\begin{figure}[htb]
    \centering
    \includegraphics[width=0.8\linewidth, trim={0.1cm 0.1cm 0.11cm 0.1cm},clip]{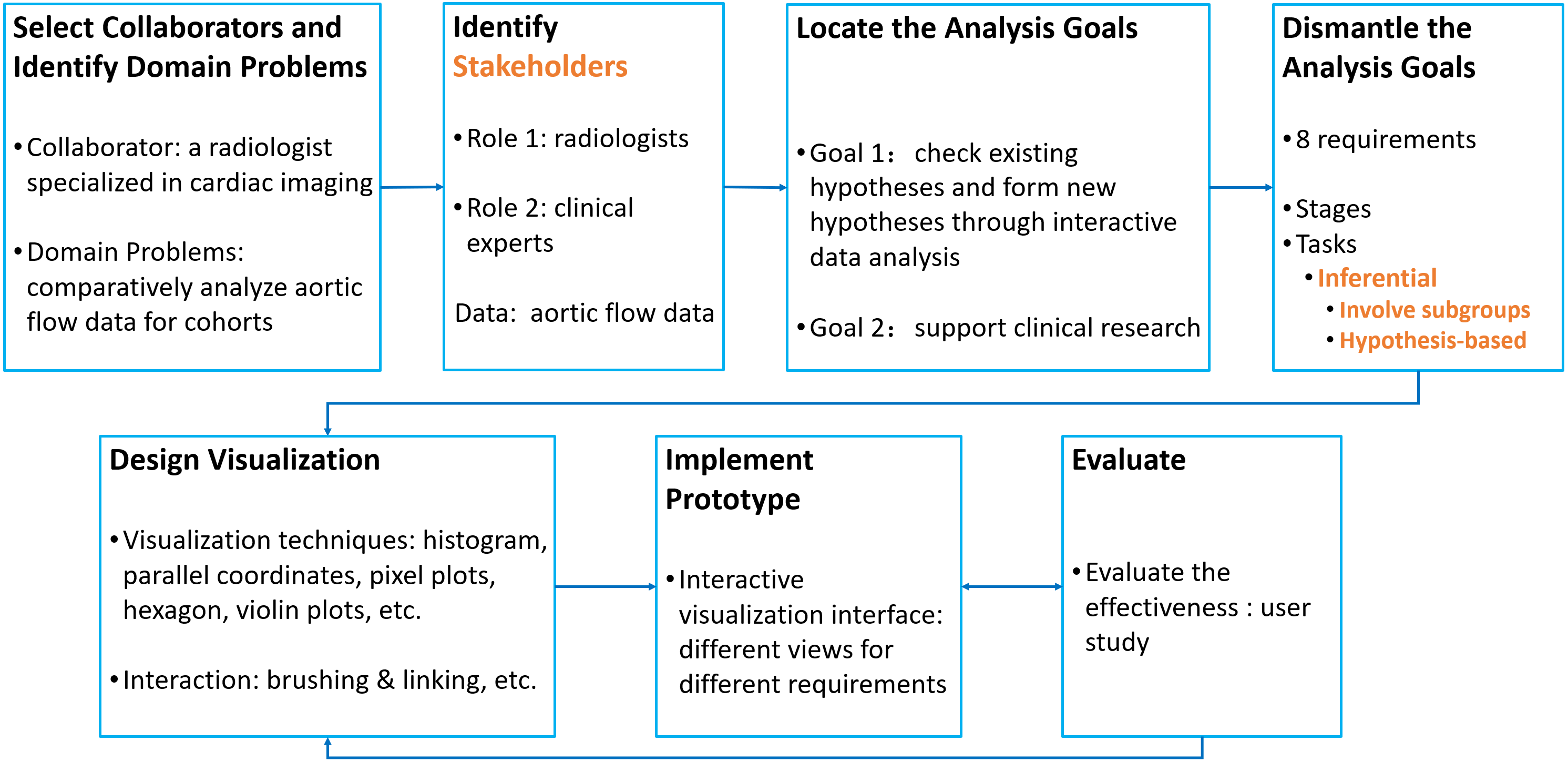}   

    \caption{The workflow of GUCCI reanalyzed with our model.}
        % \vspace{-0.5em}
    \label{fig:gucci}
\end{figure}

The fourth example reanalyzes GUCCI~\cite{Meuschke2023} (\cref{fig:gucci}), a tool for cohort-based aortic blood flow analysis. 
Researchers collaborated with an expert cardiac radiologist (30 years' experience) to define the domain problem: comparing aortic flow cohorts. 
Radiologists and clinical experts were identified as stakeholders.
Through discussions, analysis goals were established, leading to eight system requirements. 
These requirements defined specific steps (e.g., two steps for cohort visualization). 
The design incorporates diverse visualizations (e.g., histograms, parallel coordinates, pixel plots, hexagons, violin plots) and interactions (e.g., brushing \& linking). 
GUCCI was evaluated via a qualitative user study with three cardiac radiologists and two blood flow visualization experts.

\rev{
The GUCCI case highlights the importance of systematically identifying different stakeholders and integrating their needs at the early design stage. 
The tools involved two groups: radiology experts and clinical medicine experts. 
If the design only focused on the needs of radiologists, the solution might only focus on the visualization and analysis of medical imaging results, and would likely overlook the deeper needs of clinical experts - that is, using the visual results to conduct diagnostic comparisons among different patient subgroups, thereby providing support for critical clinical decisions.
Furthermore, the design of this tool fully takes into account the needs of subgroup analysis, which is consistent with the approach emphasized in our model, suggesting using a hypothesis-based method when conducting subgroup analysis. 
This method prompts researchers to think and clarify in advance the visual expression methods suitable for comparing the characteristics of different subgroups before the specific design begins, thereby ensuring that the subsequent design has stronger interpretability and specificity.
}

\subsection{Summary of the Application of the Model}
\label{sec:cases_summary}

These works were selected based on four criteria:
(1) Demonstrating dual applicability: Showcasing that the model can both \emph{reanalyze} existing works (\emph{PROACT}, \emph{e-MedLearn}, \emph{GUCCI}) and \emph{guide} new designs.
(2) Covering diverse professional backgrounds and applications: Representing key application types outlined in \cref{sec:stakeholders}. Specifically, \emph{PROACT} for patient communication, \emph{e-MedLearn} for medical student education, and \emph{GUCCI} for expert insight analysis and decision support.
(3) Covering different analytical tasks: \emph{PROACT} (descriptive), \emph{e-MedLearn} (inferential: hypothesis-free), \emph{GUCCI} (inferential: hypothesis-based).
\rev{
(4) Focusing on various data types: \emph{PROACT} (disease data encompassing diagnosis, treatment, prognosis, etc.), \emph{e-MedLearn} (various data mainly from electronic health records), \emph{GUCCI} (aortic blood flow imaging data).
}
Collectively, they demonstrate the model's applicability and generalizability from multiple perspectives.

\section{Expert Feedback on the Model}
\label{sec:stakeholder_feedback}

\rev{
We applied the proposed process model to several unpublished research projects, thereby extending its scope beyond the aforementioned use cases. 
}
\rev{
During collaborations, we proactively gathered detailed evaluations from medical experts across multiple specialties regarding the applicability of the model with three key dimensions:
(1) \textbf{Medical relevance} -- whether the emphasis of the model on distinguishing different stakeholders, stages, and subgroups accurately reflects real-world medical collaboration and analytical needs.
(2) \textbf{Effectiveness in supporting decision-making} -- the capacity of the model to assist researchers in selecting appropriate visualization techniques and evaluation methods, as well as in assessing the validity of prototype designs.
(3) \textbf{Value in cross-disciplinary collaboration} -- the value of the model in facilitating effective communication, integrating diverse perspectives, and helping interdisciplinary teams arrive at usable visualization solutions. 
}

\rev{
Medical experts unanimously agree that the proposed process model is consistent with the cognitive logic and actual needs in medicine in general. 
They point out that solving complex medical problems often requires the collaborative efforts of experts from multiple disciplines, and this model can effectively facilitate the integration of viewpoints and requirements among different stakeholders at an early stage, thereby significantly enhancing the efficiency of interdisciplinary collaboration. 
This integration not only helps to save resources but also lays a good foundation for the efficient development of subsequent analytical methods and visualization schemes. 
The phased construction method of this model is recognized as conforming to the basic logic of scientific research and clinical reasoning, supporting the gradual advancement and iterative improvement of solutions, and effectively reducing cognitive load during design and implementation. 
Clinical experts specifically note that this model clearly supports comparative analysis of different patient groups--a common requirement in actual diagnosis and treatment--thereby significantly enhancing its practicality and acceptability in clinical applications. 
A quantitative pharmacology expert L, stated that this process model is highly compatible with their actual work and can provide clear and user-friendly guidance for analytical tasks. 
A health data science expert Y, also pointed out that the model covers the entire process from requirement definition, tool design to effect evaluation, and is particularly valuable in terms of selecting visualization techniques and determining prototype interfaces, helping to further improve research efficiency.
}

\rev{
Overall, domain experts generally believe that this model can effectively promote communication among interdisciplinary teams, assist in precisely aligning and integrating diverse research needs, increase research efficiency, and thereby facilitate the construction of comprehensive and targeted visualization solutions.
}

\section{Recommendations}
\label{sec:recommendations}
Regarding the development of our new design study process model and its application in four specific examples, we put forward the following suggestions.

\textbf{Identify various stakeholders and specify target users.}
Effectively designing medical visualizations requires considering the diverse cognition and needs of users with varying medical knowledge. 
Our model integrates user-centered design~\cite{Mao2005} and participatory design~\cite{Boedker2022} principles. 
We advocate for engaging all key stakeholders throughout the design process and prioritizing the target user's perspective in visualization and prototype design.
Understanding the needs and perspectives of different stakeholders helps design various visualizations specific to varying needs, enabling solutions for a wide range of use cases~\cite{Guo2023}. 
Focusing on the target user ensures visualizations align with their background and literacy. For example, PROACT addresses potential low literacy among prostate cancer patients by utilizing simple visuals like pie charts and bar charts, balancing readability with comprehension~\cite{Hakone2017}.

\textbf{Align comprehension and cognition of different stakeholders.}
Different stakeholders may have different understandings of the same question or concept.
For example, in the work of~\cref{sec:multi_outcome}, clinical experts comprehend the interactions between different factors as correlations, while public health experts may understand them as causalities.
In addition, we learned that the causal graph in visualization can be better comprehended as a Bayesian network by medical experts, as the notion of causality is very rigorous in medicine, and, therefore, the terminology is sensitive to medical experts.
Aligning the comprehension and cognition of different stakeholders and visualization researchers through thorough discussions is a prerequisite for designing effective visualizations.

\textbf{Dismantle the analysis goals into various stages and tasks.}
Expertise barriers make medical problems inherently complex. 
According to the cognitive habits of knowledge acquisition or analytic logic, dividing goals into different stages can simplify and speed up the visualization design.
Depending on the application and target users, tasks can be categorized as inferential and descriptive, which can guide visualization design. 
For inferential tasks involving subgroup analysis, a hypothesis-based approach can improve the effectiveness of the visualization.

\textbf{Prioritize a controlled study and calculate the minimum sample size.}
We observe in our daily practice that what we do in visualization work is to design a new system or tool that stakeholders, such as medical experts, have never seen before, and this applies to many previous medical visualization works as well.
It is of significance to verify the effectiveness of the designed system to related stakeholders.
The current common practice is to collect qualitative user comments and quantitative results on the designed system through user evaluations, controlled studies, or case studies, usually with relatively few participants (fewer than a few dozen). 
Although different assessment methods provide complementary insights, medical experts are more convinced by the results of controlled studies.
Therefore, we recommend prioritizing controlled experiments for evaluating medical visualization systems. 
To address the limitations of small sample sizes that threaten universality, reliability and effectiveness, we further advocate conducting pilot studies to determine the minimum sample size and then recruit this minimum sample size in formal controlled studies.

However, we also realize that recruiting participants is a challenge, especially for applications involving medical professionals. 
For non-professionals (e.g., patients, students), crowdsourcing can be an actionable option.  
For expert recruitment, we recommend prioritizing meeting minimum sample sizes. 
If still unachievable, consider supplementing with robust quantitative evaluations. 
In addition, ethical approval should always be ensured before conducting research involving humans to ensure compliance and prevent delays in the study.

Researchers in the interdisciplinary field of medical visualization in the future need to balance the reliability and feasibility of the evaluation methods.

\section{Discussion}
\label{sec:discussion}
In this section, we first reflect on the proposed design study process model, and then we compare it to existing models.

\subsection{Reflection on the New Process Model}
\label{sec:reflection}
The model is based on the analysis and summary of medical-related visualization papers (\cref{sec:features}), references on other previous models (\cref{sec:models}), and our own interdisciplinary research experience. 
Although the year range of the considered papers is limited, these papers cover various data types, applications, stakeholders, etc.. 
The model has the potential to be applied to many medical visualization and visual analysis works, as illustrated by the examples mentioned in~\cref{sec:examples}.

The main difference between the proposed model for medical visualization and the previous design study process models is that it emphasizes the importance of distinguishing different stakeholders before the specific design begins. 
This point of view stems from the finding that 62 of the 78 reviewed papers involve different stakeholders, and that the design or evaluation of a particular system or tool in these 62 papers took into account the needs and backgrounds of various individuals.
This is consistent with previous studies stating that user and environment characteristics affect the type of visualization tools needed~\cite{Crisan2021} and the effect of interaction techniques adopted~\cite{Myers2019}.
Compared to considering only individual stakeholders, considering different stakeholders can help obtain more comprehensive and differentiated requirements before the design, thereby facilitating the development of a universal and targeted visualization design and avoiding potential iterations. 
However, in cases where stakeholder demands are highly dispersed or complex, this may pose challenges for visualization design, such as requiring more time and resources for consultations, or making it difficult to integrate these requirements. 
When it is difficult to reconcile multiple demands, prioritizing the needs of the target users may be a reasonable approach.

Dismantling goals into different stages is the second emphasis of this proposed model.
We believe that distinguishing between different stages according to the analysis process or knowledge acquisition habits will benefit the design study of future medical visualization work.
Some medical visualization and visual analysis work focuses on comparing subgroups, and hypothesis-based analysis helps design suitable visualizations to assist in analyzing the differences between different subgroups. 
This forms the third emphasis of the proposed model -- dividing inferential tasks into hypothesis-based and hypothesis-free according to whether subgroups are involved.

Although the model is proposed based on the summary and analysis of research work on medical visualization, other fields may also benefit from this model.
For example, considering multiple stakeholders helps achieve a more comprehensive visualization design, and differentiating stages according to the analysis logic facilitates a logically reasonable and progressive design of visualization solutions.

\subsection{Comparison to Other Process Models}
\label{sec:model_comparison}
Our model can relate to some previous process models. 
In the following, we discuss the connections and differences of our model mainly to the well-known nested model~\cite{Munzner2009} and the nine-stage framework~\cite{Sedlmair2012}.
\begin{itemize}
    \item  \textbf{Collaborator selection} is similar to \emph{winnow} in the nine-stage framework~\cite{Sedlmair2012}. 
    The difference is that we focus specifically on collaborating with medical experts and provide criteria for their selection.
    The aim is to clarify whose specific problems and needs can be addressed by visualization.
    \item \textbf{Domain problem identification} is similar to the first level of \emph{domain problem characterization} mentioned in the nested model~\cite{Munzner2009} and \emph{discover} in the nine-stage framework~\cite{Sedlmair2012}, which also calls for consistent understanding of various concepts by people with different disciplinary backgrounds (e.g., visualization designers and medical experts), to make sure that the specific medical problem is correctly comprehended by all parties.
    \item \textbf{Stakeholders identification} is a unique step of our model. 
    \rev{
    Conducting a systematic analysis of stakeholders at the early stage of the project can effectively identify the core demands and key challenges of all parties, providing a clear direction for subsequent visual design and avoiding repeated revisions and resource waste caused by unclear goals. 
    By inviting stakeholders from various backgrounds to participate in in-depth discussions at the very beginning of the design process, it is possible to build consensus and integrate diverse needs. 
    This enables the final visualization solution to break through the limitations of a single function and become a collaborative link that connects different professional fields and serves a common goal~\cite{Mistelbauer2023}. 
    As a result, the depth of understanding and the breadth of application of the solution can be significantly enhanced.
    }
    \item \textbf{Analytical goals localization} shares some similarities with \emph{domain problem characterization} in the nested model~\cite{Munzner2009} and \emph{discover} in the nine-stage framework~\cite{Sedlmair2012}. 
    The difference lies in that the previous models mix questions, objectives, tasks, data, and other elements, whereas in this model, the step focuses on first obtaining the comprehensive analysis objectives from different stakeholders to support subsequent effective visualization design.
    \item \textbf{Goals dismantling} is similar to \emph{operation abstraction} in the nested model~\cite{Munzner2009} and \emph{discover} in the nine-stage framework~\cite{Sedlmair2012}, and the aim is to figure out a set of design requirements in the vocabulary of computer science, guiding visualization researchers in the design. 
    The common practice of previous models is to abstract goals to different tasks and data types~\cite{Munzner2009}.
    We add that goals can be broken down into different stages according to analytic logic or cognitive habits. 
    Additionally, we propose taking an inferential or descriptive approach to analyze the task, depending on the target users and applications.
    We also suggest applying hypothesis testing to help with inferential task analysis if it involves subgroups.
    This step can simplify and speed up the visualization design. 
    \item \textbf{Visualization design} integrates both \emph{encoding/interaction technique} and \emph{algorithm} design in the nested model~\cite{Munzner2009}.
    While aligning with the \emph{design} in the nine-stage framework~\cite{Sedlmair2012}, our model refines this by relocating \emph{data abstraction} to the earlier \emph{goals dismantling} phase, and centering design around differentiated requirements.
    We further enhance this stage by providing practical guidance on selecting visualization techniques.
    \item \textbf{Prototype deployment} is similar to \emph{implement} in the nine-stage framework~\cite{Sedlmair2012}.
    In this step, we focus on the implementation of the prototype and not on \emph{usability}. 
    In addition, we suggest that different interfaces can be designed for different stages or stakeholders.
    \item \textbf{Evaluation} integrates \emph{implement}, \emph{deploy},  and \emph{reflect} of the nine-stage framework~\cite{Sedlmair2012}.
    We also provide some evaluation methods that future researchers can refer to.
    In addition, we propose evaluation methods that are more recognized by medical experts according to the specific characteristics of the medical field, which also forms one of our recommendations. 
\end{itemize}

The proposed model tailors visualization design to medical contexts through domain-aware refinements.
It expands on certain steps from the previous model; for example, the \emph{discover} stage in the nine-stage framework corresponds to three steps in our models -- \emph{domain problem identification}, \emph{analytical goals localization}, and \emph{goals dismantling}, which can enhance the model's operability. 
Embedded step-by-step guidance enables researchers to systematically develop effective medical visualization solutions.

\section{Conclusion}
We have introduced a design study process model for medical visualization.
The model is formulated based on characteristics of medical data analysis problems and tasks summarized by a literature review, combined with our own interdisciplinary research experience, and referring to previous models.
It features three factors: \emph{stakeholders}, \emph{stages}, and \emph{subgroup analysis} that need to be carefully considered in a design study for medical visualization.
With four use cases of medical-related visualization works, we demonstrate the usefulness of the model. 
We provide recommendations for each step of the model.
In a discussion, we reflect on the model and delineate it from existing models.

%% if specified like this the section will be omitted in review mode
\acknowledgments{%
    This work was supported by the National Natural Science Foundation of China (grant 62372012).%
}

\bibliographystyle{abbrv-doi-hyperref}

\bibliography{designStudyModel}

\appendix % You can use the `hideappendix` class option to skip everything after \appendix

% \section{About Appendices}
% Refer to \cref{sec:appendices_inst} for instructions regarding appendices.

% \section{Troubleshooting}
% \label{appendix:troubleshooting}

% \subsection{ifpdf error}

% If you receive compilation errors along the lines of \texttt{Package ifpdf Error: Name clash, \textbackslash ifpdf is already defined} then please add a new line \verb|\let\ifpdf\relax| right after the \verb|\documentclass[journal]{vgtc}| call.
% Note that your error is due to packages you use that define \verb|\ifpdf| which is obsolete (the result is that \verb|\ifpdf| is defined twice); these packages should be changed to use \verb|ifpdf| package instead.

% \subsection{\texttt{pdfendlink} error}

% Occasionally (for some \LaTeX\ distributions) this hyper-linked bib\TeX\ style may lead to \textbf{compilation errors} (\texttt{pdfendlink ended up in different nesting level ...}) if a reference entry is broken across two pages (due to a bug in \verb|hyperref|).
% In this case, make sure you have the latest version of the \verb|hyperref| package (i.e.\ update your \LaTeX\ installation/packages) or, alternatively, revert back to \verb|\bibliographystyle{abbrv-doi}| (at the expense of removing hyperlinks from the bibliography) and try \verb|\bibliographystyle{abbrv-doi-hyperref}| again after some more editing.

\end{document}